# Mechanical Effects of Optical Vortices[1]

**N.R. Heckenberg, M.E.J. Friese, T.A. Nieminen and H. Rubinsztein-Dunlop**

*Centre for Laser Science, Department of Physics, The University of Queensland,*

*Brisbane QLD 4072, Australia*

## Introduction

As has been pointed out by Berry[1], the concept of phase is an abstract one, and the idea of a singularity perhaps more so, so that a singularity in phase is a double abstraction. Besides, even the fact that light has an oscillating or wave nature at all is not often obvious. It is therefore particularly gratifying that optical phase singularities, which at first were experienced only rather negatively as dark spots or indirectly through defects in interference patterns, have in recent times been shown to be tangible enough to exert mechanical forces capable of rotating and moving small objects about. Of course, the forces are exerted by the light field surrounding the singularity but their characteristics are determined by the characteristic radial increase in irradiance and by the helical nature of the wavefronts associated with the singularity (the optical vortex), which can therefore be held responsible for the effects.

The fact that light carries energy, momentum, and angular momentum has been well understood since the work of Maxwell (1873) but the small sizes of the associated forces, together with the interference of spurious effects such as radiometer, thermophoretic and photophoretic effects in low pressure gases, made direct measurements difficult. Lebedev[2] and Nichols and Hull[3] were probably the first to make measurements of radiation pressure, while in a well-known experiment, Beth[4] measured the angular momentum carried by circularly polarized light.

---

[1] N. R. Heckenberg, M. E. J. Friese, T.A. Nieminen and H. Rubinsztein-Dunlop, "Mechanical effects of optical vortices", pp 75-105 in M. Vasnetsov (ed) *Optical Vortices* (*Horizons in World Physics* **228**), Nova Science Publishers (1999)

Recently a microscopic parallel of Beth's experiment was carried out. In this experiment, the angular momentum carried by circularly-polarized light in a Gaussian beam was transferred to a microscopic fragment of birefringent material causing it to rotate[5].

The issue of the mechanical effects of optical vortices was first raised in 1992 by Allen et al.[6] . They pointed out that Laguerre-Gaussian (LG) laser modes carry angular momentum associated with their optical vortices and proposed an experiment reminiscent of that of Beth to observe transfer of the vortex angular momentum to a macroscopic object. On the basis of classical electromagnetic theory, they showed that the ratio of angular momentum flux to energy flux for a LG mode of order *l* can be written as

$$\frac{\mathbf{J}}{c\mathbf{P}} = \frac{(l+\mathbf{s}_z)}{\mathbf{w}} \qquad (1)$$

where $\mathbf{s}_z = \pm 1$ for left-handed or right-handed circularly polarized light and $\mathbf{s}_z = 0$ for linearly polarized light, and $\mathbf{P}$ is the linear momentum density. This can be conveniently interpreted as each photon of energy $\hbar\mathbf{w}$ carrying $\mathbf{s}_z$ quanta of "spin" angular momentum associated with the polarization state and *l* quanta of "orbital" angular momentum associated with the spatial distribution of the field, each quantum having magnitude $\hbar$.

Abramochken and Volostrukov[7] had shown that LG modes could be converted to Hermite-Gaussian and back by passing through mode converters of two cylindrical lenses. Such converters have been well studied and have since found several uses[8,9,10].

In the experiment proposed by Allen et al.[11], which was to show that a laser mode carries angular momentum associated with its optical vortex, a pair of cylindrical lenses suspended on a torsion fibre would reverse the helicity of a LG beam and suffer a reaction torque. However, this torque is very small for reasonable power beams and it has turned out to be easier to measure such effects on a more microscopic scale in the context of optical trapping experiments as discussed below.

Optical trapping is achieved through forces arising from exchange of momentum between incident light and the trapped object. The local flow of momentum in the light field is in the direction of the local Poynting vector and hence perpendicular to the wavefronts. Since optical vortices have wavefronts which are helical and therefore inclined to the axis of propagation of the beam they can apply circumferential forces and hence torques about the beam axis. It is in this way that angular momentum is transferred to some object from the beam.

The LG modes are only paraxial solutions of the wave-equation, but optical trapping experiments are carried out with strongly focussed, and therefore non-paraxial, light beams. In this case it is not in principle permissible to separate the angular momentum of a beam into independent orbital and spin components. This problem has been considered by Barnett and Allen[12] who have derived a correction term which is fortunately negligibly small in most cases of interest.

Other evidence for vortex angular momentum comes from experiments where LG beams have been used in second harmonic generation[13,9], in which LG beams of twice the order, corresponding to twice the angular momentum per photon were produced. This is consistent with a picture where two photons combine to form one with twice the energy, twice the momentum, and twice the angular momentum, satisfying the conservation law for each quantity.

In this chapter we concentrate on the forces and torques exerted on transparent and absorbing particles trapped in laser beams containing optical vortices. We review previous theoretical and experimental work and then present new calculations of the effect of vortex beams on absorbing particles.

## 1. Optical vortices in trapping experiments

### 1.1 The use of optical vortex modes to trap high refractive index transparent particles

As early as 1974, Ashkin and Dziedzic[14] showed that a $TEM^*_{01}$ laser mode could be used to stably levitate various types of spheres. Although there is no guarantee that the mode was not an incoherent superposition of $TEM_{01}$ and $TEM_{10}$ modes with different frequencies and hence not strictly a vortex, it might have been, at least some of the time. Their paper describes levitation of solid transparent spheres in a single vertically upward propagating doughnut beam, and levitation of hollow transparent and metallic spheres using a combination of two counter-propagating beams and one vertically upward beam.

However, until the advent of the single-beam gradient optical trap[15], manipulation of microscopic objects using laser beams was not a truly practicable reality.

In a single-beam optical gradient trap, one laser beam is focussed in such a way as to produce a 3-dimensional trapping potential. In most commonly used experimental set-ups for a single-beam optical trap, the laser beam is directed through a high numerical aperture (NA ≈ 1.3) 100× objective lens. The beam is shaped in such a way as to ensure filling of the back pupil of the objective lens in order to achieve the tightest possible focus at (or near) the specimen plane. High numerical aperture is very important as it enables maximization of the light intensity gradient near the focal plane, providing the most stable trapping in the axial direction.

The minimum focal spot size (of the order of 1-2 $\mu$m) and maximum intensity gradient result in the strongest trapping force on an object to be trapped and manipulated. An object to be trapped is usually suspended in an appropriate solution, a drop of which is placed between a microscope slide and coverslip. Imaging of the particle and trap is most often performed through the same lens that is used to bring the beam to a focus, either through microscope eyepieces or a CCD camera. A variety of lasers has been used for trapping, such as He-Ne, CW Nd-YAG and Ti-Sapphire lasers with powers ranging from a few to hundreds of mW.

Often the laser beam is directed into a commercially available microscope and only minor modifications are needed in order to perform laser micromanipulation. In most experiments, the trapped specimen has to be moved with respect to the rest of the sample. This can be achieved either by moving the trap itself or by moving the specimen, the simplest option being to move the microscope stage, causing the entire sample to move while the trapped particle remains stationary with respect to the objective. More complicated options include setups where the direction of the beam entering the back pupil of the objective is controlled, allowing the trap to be moved while the sample remains stationary.

The efficiency of the trap can be most simply described by a radial and an axial trapping efficiency, which are commonly measured by determining the maximum velocity at which the particle can be dragged through the surrounding fluid and still remain trapped by a given laser output power, or by measuring the effective force constants. Radial trapping efficiency is in general higher than axial efficiency (2 - 3 times greater for glass spheres 1 - 10 $\mu$m in diameter[16]).

Wright et al.[17] were the first to model the optical tweezers trapping efficiency using a ray optics approach. A similar model was used by Ashkin[18], who showed that within a ray optics model, rays with a high convergence angle (around 70°) contribute more to the trapping force than those from the center of the beam. This leads to the postulate that a beam with a central dark spot (for example a $TEM^*_{01}$ doughnut beam) may provide more efficient trapping. Usually the "scattering force" acting downwards tends to push the particle away from the beam focus, counteracting the trapping. A beam with a central dark spot reduces this force, improving axial trapping.

Ashkin's investigation of trapping efficiency for $TEM_{00}$ and $TEM^*_{01}$ laser beams and particles large compared to the wavelength showed that a $TEM^*_{01}$ beam achieves better on-axis trapping (around 25% better than a Gaussian beam of the same spot-size having the same power), but the radial trapping efficiency of a doughnut beam is calculated to be around 10% less than that of an equivalent Gaussian beam.

As optical tweezers are primarily used to manipulate live biological materials, achieving the most efficient trapping system has long been a concern, to minimize the possibility of damage to the specimen. Recent work by Friese at al.[19] and Simpson et al.[20] has shown that the prediction of greater axial trapping efficiency using $TEM^*_{01}$ beams is indeed correct, and that by using Laguerre-Gaussian modes of various orders, axial trapping efficiency can be increased by several times[21]. However, the predicted decrease in radial trapping efficiency is to our knowledge yet to be verified.

## 1.2    The use of LG modes to trap non-transparent or low-index particles

Although the use of optical vortices in biological and biophysics research is as yet uncommon, doughnut beams and other hollow beam arrangements have been widely used to manipulate non-transparent (both absorptive and reflective) and low-index (refractive index lower than their surrounding medium) particles. The early work of Ashkin and Dziedzic[14] and Roosen and Imbert[22], who manipulated reflective particles using counter-propagating doughnut beams, was followed by experiments with various types of single-beam optical trap. These traps all use an arrangement similar to the single-beam gradient trap: a single laser beam is focussed to a waist on the order of one wavelength across using a high NA microscope objective. Axial forces due to absorption or back reflection of light preclude 3-dimensional trapping, but particles can be confined 2-dimensionally against a substrate like a microscope slide. These trapped objects can then be manipulated in one plane only, as opposed to the 3-dimensional trapping effect of the single-beam gradient trap.

Beams with central intensity minima have been employed to trap non-transparent and low-index particles because simple theory predicts that these particles cannot be trapped in the usual Gaussian beam. The same ray optics model used to explain high-index particle trapping shows that low-index particles are repelled from intensity maxima, and a similar theory was used to show that reflective particles would suffer similar effects[23]. Experiments showed that strongly absorbing particles could also be trapped in a doughnut beam[24]. It has since been shown theoretically and practically that both reflective[25] and absorptive[26,27] particles can in fact be trapped 2-dimensionally without a central dark spot (although trapping is less stable than in a doughnut beam), and also that some very small metallic particles around 10 nm in size can even be trapped 3-dimensionally in a single-beam gradient trap[28] as they are small enough to experience some induced polarization effects. However, doughnut modes are by far the most commonly used beams for manipulation of these types of particles, although other arrangements such as the circular scanning of a laser beam around a particle[23] have also been successful. Low-index particles (2-dimensional trapping of bubbles[29] and 3-dimensional trapping of hollow glass spheres[30]), metal particles[25,24] and absorbing particles[31,32,33] have also been trapped using $TEM^*_{01}$ laser beams. However, in most cases, the reason for employing these modes has been to utilize the central intensity minimum, rather than any other interesting properties of the beam. Recently though there has been an increasing interest in using optical traps to investigate these properties, and to observe their effects on microscopic particles.

### 1.3  Trapping absorbing particles with LG modes

The first experimental evidence of the mechanical effects of the orbital angular momentum of LG modes was the optically induced rotation of black ceramic particles of around 1 - 2 $\mu$m diameter, while they were trapped in a highly focussed doughnut beam produced using computer-generated holograms[31]. The main evidence cited in this work was the dependence of the sense of rotation of the particles on the sense of the helicity of the vortex. In a similar experiment[32] a Dove prism was included in the optical path. The removal of the Dove prism reversed the helicity of the beam: the purpose of this arrangement was to observe the sense of rotation of a given particle as the helicity changed direction. A reversal of the rotation direction was observed, further evidence that the particle rotation was caused by the optical vortex.

Other experiments along similar lines have demonstrated the mechanical equivalence of optical angular momentum associated with polarization and optical vortices. As was mentioned in the introduction, Barnett and Allen[12] calculated that a polarized charge $l$ Laguerre-Gaussian beam carries angular momentum $(l + s_z)\hbar$ per photon in the paraxial limit (where $s_z$ is ±1 and 0 for circularly- and plane-polarized light respectively). Friese et al.[33] showed that an absorbing particle rotating with frequency $f$ while trapped in a charge 3 Laguerre-Gaussian mode speeds up to a frequency $4/3f$ when the beam is made circularly-polarized with the same sense as the vortex, and slows down to $2/3f$ when the beam is circularly-polarized the opposite way.

A similar experiment was reported by Simpson et al.[20] using a charge one doughnut beam, showing that the torque due to the beam helicity could be cancelled out by the torque due to circular polarization of the opposite sense.

## 2. Calculation of optical forces

The forces exerted on particles by a laser beam depend on the properties of both the beam and the particle. A number of distinct mechanisms act when light is incident on particles of different types, such as gradient forces acting on transparent particles and atoms, and the transfer of momentum through absorption of photons from the beam. These different types of particles will be considered separately.

While a non-paraxial beam is necessary for effective laser trapping, the use of the paraxial approximation is convenient, and does not seem to introduce excessive error. Errors due to aberration within the optical system are expected to be greater.

## 2.1 Description of the laser beam

The most common type of phase singular laser beam used for trapping is a Laguerre-Gaussian doughnut beam $LG_{pl}$ (described by a radial mode index $p$ and an azimuthal mode index, or charge, $l$). The radial mode index $p$ is equal to zero for doughnut beams used for trapping. Beams with azimuthal mode indices of $l = 1$ to $l = 3$ are most common. Due to the cylindrical symmetry of the beam, it is convenient to use a cylindrical coordinate system, with radial, azimuthal and axial coordinates $r$, $f$ and $z$ and corresponding unit vectors . In this coordinate system, the linear momentum flux of a $LG_{pl}$ laser beam is given by the time-averaged Poynting vector, **S**,

$$\mathbf{S} = \frac{ce}{2} E_0^2 \left( \frac{zr}{z_r^2 + z^2} \hat{\mathbf{r}} + \frac{l}{kr} \hat{\mathbf{f}} + \hat{\mathbf{z}} \right) \quad (2)$$

where $z_r$ is the Rayleigh range, $k$ is the wavenumber of the beam and $E_0$ is the amplitude of the beam, given by

$$E_0 = \sqrt{\frac{2}{ce}} \sqrt{\frac{2p!P}{p(l+p)!}} \left( \frac{r\sqrt{2}}{w(z)} \right)^l L_p^l \left( \frac{2r^2}{w^2(z)} \right) \frac{1}{w(z)} e^{\frac{-r^2}{w^2(z)}} \quad (3)$$

where $P$ is the beam power, $w(z)$ is the beam width, $L_p^l$ is the generalized Laguerre polynomial, $c$ is the speed of light and $e$ is the permittivity. The Rayleigh range $z_r$ and the beam width $w(z)$ are related to each other and to the beam waist spot size $w_0$ by the standard relations:

$$w^2(z) = \frac{2(z_r^2 + z^2)}{kz_r} \quad (4)$$

$$w(z) = w_0 \sqrt{1 + \frac{z^2}{z_r^2}} \quad (5)$$

$$z_r = \frac{kw_0^2}{2} \quad (6)$$

## 2.2 Transparent particles

The trapping of transparent particles can be well understood in terms of gradient (or refractive or polarization) forces and scattering (including reflection and absorption) forces. The gradient forces are proportional to the intensity gradient, while the scattering forces are proportional to the intensity (thus the importance of having a highly convergent beam). These forces depend strongly on the composition of the trapped particle and the surrounding medium.

The gradient force is due to the spatial variation of the interaction energy between the electric field of the incident beam and the induced polarization due to this field within the particle. The dipole moment per unit volume, **d** in an isotropic particle induced by an applied electric field **E** is

$$\mathbf{d} = \frac{\varepsilon_0 (1-n^2)}{n^2} \mathbf{E} \tag{7}$$

where $n$ is the refractive index of the particle. The interaction energy density $U$ is

$$U = -\mathbf{E} \cdot \mathbf{d} = \varepsilon_0 \left( \frac{(1-n_m^2)}{n_m^2} - \frac{(1-n^2)}{n^2} \right) E^2 \tag{8}$$

where $n_m$ is the refractive index of the surrounding medium, giving a time-averaged force per unit volume of

$$\mathbf{F}_G = \frac{\varepsilon_0}{2} \left( \frac{1}{n_m^2} - \frac{1}{n^2} \right) \nabla E^2 \tag{9}$$

For a small particle, with an approximately constant time-averaged electric field, calculation of the total force is straightforward, but for a larger particle, which will affect the intensity distribution of the beam, the case is more complicated.

The total force acting on a particle can be written in terms of a macroscopic polarizability $\alpha$ and an absorption cross-section $\sigma$ as

$$\mathbf{F} = -\frac{1}{2} n_m \alpha \nabla E^2 + \frac{\sigma}{c} \mathbf{S} \tag{10}$$

Provided that the gradient force is sufficient to overcome the scattering forces, the particle is attracted towards the highest intensity regions of the beam. As a result, the equilibrium position of particles trapped in a Laguerre-Gaussian beam depends on the relative size of the beam and the particles (see figure 1).

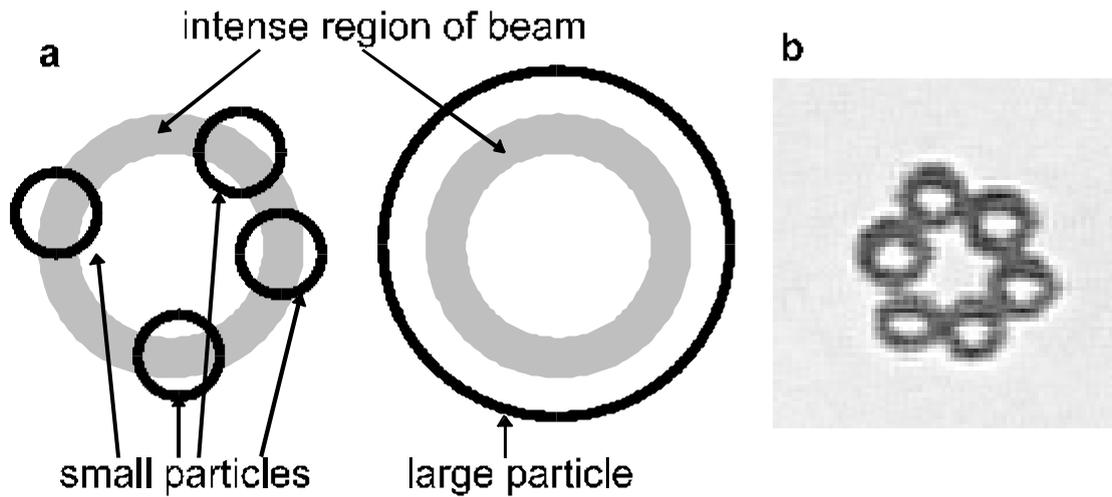

**Figure 1.** Trapping positions for transparent particles. **(a)** Trapped transparent particles are attracted to the regions of high intensity. Small particles are trapped in the high intensity ring, while particles larger than the beam will be trapped on the beam axis. **(b)** A photograph of 6 10 *m*m spheres trapped in a doughnut beam, courtesy of Gahagan and Swartzlander[30].

## 2.3 Absorbing and reflective particles

Absorbing and reflective particles are trapped (two-dimensionally) through the transfer of momentum from the beam; as photons are absorbed from the beam or reflected by a particle, momentum, angular momentum and energy are transferred from the beam to the particle. The forces acting on such particles can be calculated readily, as the difference between the incident and outgoing momenta is transferred to the particle.

Development of a useful general model for calculation of forces acting on reflective particles presents some difficulty, as the geometry of the particle affects the directions into which reflection occurs. For clarity, only the simpler case of absorbing particles will be considered in detail below. If a particle is highly absorbing, all of the energy and momentum carried by the portion of the beam intercepted by the particle will be absorbed. The resultant optical forces depend less on the optical properties and shape of the particle than when transparent or reflective particles are trapped. The primary consideration is the geometry of the beam.

### 2.3.1 Microscopic transfer of momentum and energy

The transfer of momentum from the laser beam to an absorbing particle can be readily understood in terms of the absorption of photons from the beam. A small area element **dA** of a highly absorbing particle experiences a force (i.e. a rate of momentum transfer) given by the Poynting vector:

$$\mathbf{F}_{absorbing} = \frac{1}{c}\mathbf{S}\cdot(-\mathbf{dA})\frac{\mathbf{S}}{|\mathbf{S}|} \qquad (11)$$

Reflective particles will redirect rather than absorb this momentum, so the force acting on a reflective particle will be

$$\mathbf{F}_{reflective} = \frac{-2(\mathbf{S}\cdot\mathbf{dA})^2\mathbf{dA}}{c|\mathbf{S}||\mathbf{dA}|^2} \qquad (12)$$

In this case, the direction of the resultant force depends on the orientation of the surface. An absorbing area element **dA** will also absorb angular momentum if the beam is circularly polarized, with each photon having angular momentum of magnitude $\hbar$. The rate of absorption of angular momentum by a small section of the particle is given in terms of the Poynting vector **S** and the wavenumber $k$ by

$$\mathbf{t} = \frac{s_z}{k}\mathbf{S}\cdot(-\mathbf{dA})\frac{\mathbf{S}}{|\mathbf{S}|} \qquad (13)$$

where $s_z = \pm 1$ for left- and right-circular polarization and $s_z = 0$ for plane polarization. The rate of absorption of energy can also be written in terms of the Poynting vector as

$$P_a = c\mathbf{S}\cdot\mathbf{dA} \qquad (14)$$

where $c$ is the speed of light in the medium.

### 2.4 Macroscopic effects

It will usually be necessary to integrate the Poynting vector over the surface of the absorbing particle in order to extend this microscopic momentum and energy transfer to the particle as a whole. The calculation can be readily performed numerically by dividing the particle into a number of small area elements **dA** over each of which the Poynting vector is approximately constant, and integrating equation 11. Only the illuminated surface of the particle (i.e. where the product **S**·(-**dA**) < 0) needs to be considered. A useful approximation is to simply represent the particle as a disk with the same cross-section, essentially applying the paraxial approximation to the illumination of the particle (note that this cannot be applied to most reflective particles).

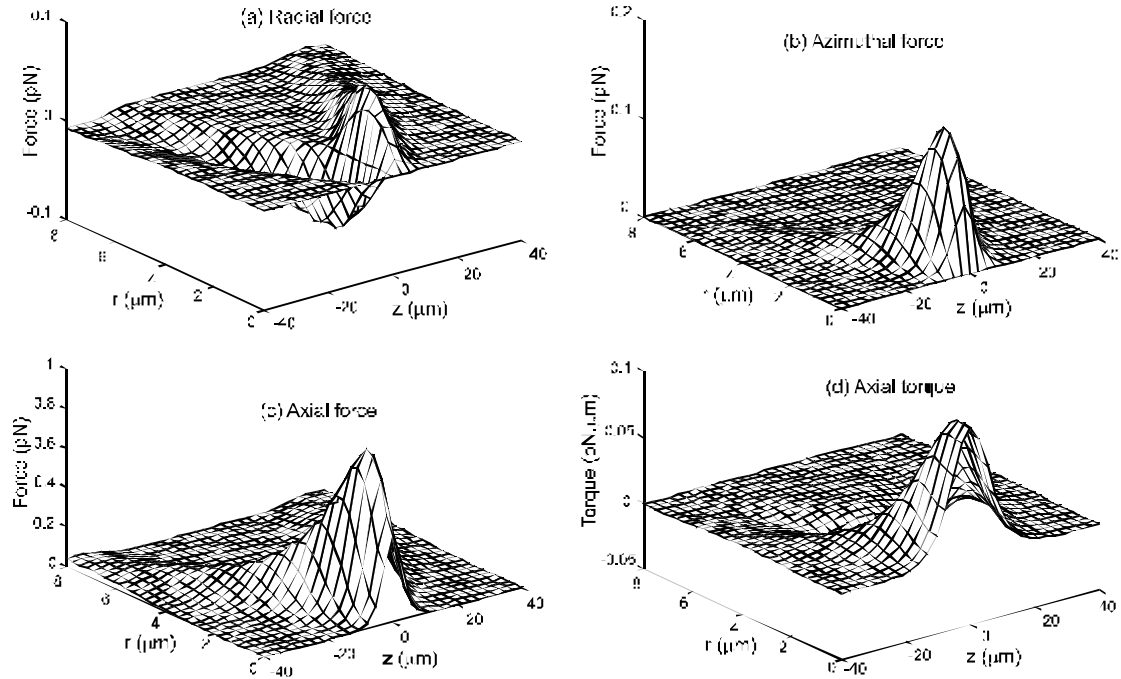

**Figure 2.** Force and torque due to a laser beam. The forces and torque acting on an absorbing particle of radius 1 $m$m trapped in water by a plane-polarized $LG_{03}$ laser beam of power 1 mW, wavelength 632.8 nm and waist width $w_0 = 1$ $m$m are shown. The beam is propagating in the $+z$ direction, with the focal plane at $z = 0$. The radial force is responsible for the trapping of the particle (which can only be trapped in the region where the radial force is negative i.e. restoring) while the azimuthal force causes orbital motion of the particle about the centre of the trap. As the beam in this case is plane-polarized, the torque about the beam axis is due to the azimuthal force only, while for a circularly-polarized beam, the torque due to absorption of angular momentum carried by the photon spin is also important.

It can be seen that the transfer of angular momentum due to polarization results in a torque acting on the particle. The transfer of linear momentum results in a force acting on the particle, and in most cases, a torque as well. The distinction between spin angular momentum due to the polarization and orbital angular momentum due to the helical structure of the beam is clearly shown by the differences in the ways in which they act on the particle.

## 2.4.1 Trapping

Equilibrium points in these force fields exist only where the total force is zero, and a particle can be trapped only at a stable equilibrium point. Thus, it can be readily seen that an absorbing particle cannot be axially trapped without an external force (such as gravity, viscous drag due to convection, or a reaction force due to the particle resting on the bottom of the trapping cell) acting on it.

The particle can be trapped radially in the portion of the beam which is converging towards the beam waist. The particle cannot be trapped after the beam waist, as the convergence of the beam required for trapping no longer exists. The absorbing particle trap can therefore be considered to be a two-dimensional trap.

The trapping of reflective particles depends on the shape of the particle as the direction of optical force depends on the direction of the normal to the surface of the particle. Spherical particles, with the normal of the illuminated surface directed away from the centre of the trap, will be trapped. Disk-shaped particles, with the normal for most of the surface directed parallel to the beam axis, will only experience a trapping force due to reflection from the edges (see figure 3). The trapping in this case will be strongly dependent on the thickness of the particle, and will occur before the focus as in the case of absorbing particles. The force component responsible for trapping is denoted in the figure by $\mathbf{F}_r$.

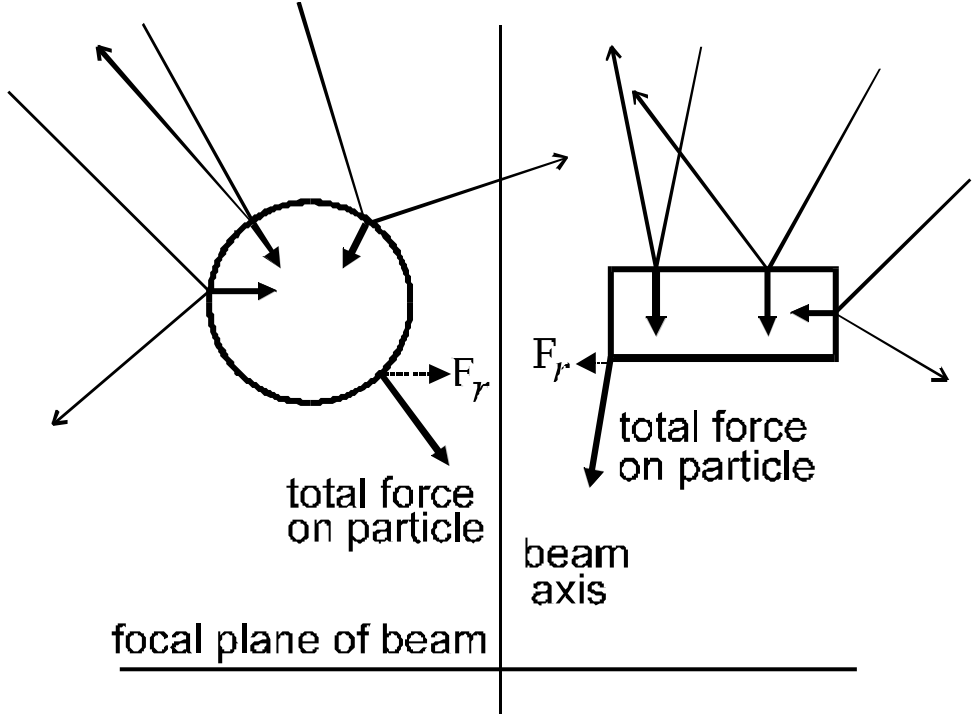

**Figure 3.** Trapability of reflective particles. The force acting on a reflective particle depends on its shape. If the total force does not include an inward radial component, the particle will not be trapped.

### 2.4.2 Particles trapped on the beam axis

The total torque due to polarization on a particle of radius *a* of absorptivity **a** trapped on the beam axis is given by

$$\boldsymbol{t}_p = \frac{\boldsymbol{a}\boldsymbol{s}_z P}{\boldsymbol{w}} \left( 1 - e^{\frac{-2a^2}{w^2(z)}} \sum_{k=0}^{l} \frac{1}{k!} \left( \frac{2a^2}{w^2(z)} \right)^k \right) \quad (15)$$

where **w** is the angular frequency of the light. The term in the brackets is the fraction of the beam power intercepted by the particle. The total torque can be found by combining this with the torque due to the orbital angular momentum carried by the helicity of the beam

$$\boldsymbol{t}_o = \frac{\boldsymbol{a}\,l\,P}{\boldsymbol{w}} \left( 1 - e^{\frac{-2a^2}{w^2(z)}} \sum_{k=0}^{l} \frac{1}{k!} \left( \frac{2a^2}{w^2(z)} \right)^k \right) \quad (16)$$

As the particle is trapped, the torque acting on it will cause the particle to rotate on the spot. The terminal angular velocity $\Omega$ in a viscous medium will depend on the drag torque, which for a smooth spherical particle of radius *a* in a medium of viscosity **h** is given by

$$t_d = -8\pi\eta a^3 \Omega \tag{17}$$

and will equal the optical torque when the terminal angular velocity is reached. The resultant spin rates for particles trapped on the beam axis are

$$\Omega = \frac{\alpha(s_z + l)P}{8\pi\eta a^3 \omega}\left(1 - e^{\frac{-2a^2}{w^2(z)}}\sum_{k=0}^{l}\frac{1}{k!}\left(\frac{2a^2}{w^2(z)}\right)^k\right) \tag{18}$$

Figure 4 shows spin rates calculated for typical experiments where rotation rates of a few Hz are observed. As the particle size is increased, more angular momentum is absorbed until the particle is larger than the beam. Any further increase in radius increases the drag without increasing the optical torque, leading to a decrease in rotation speed.

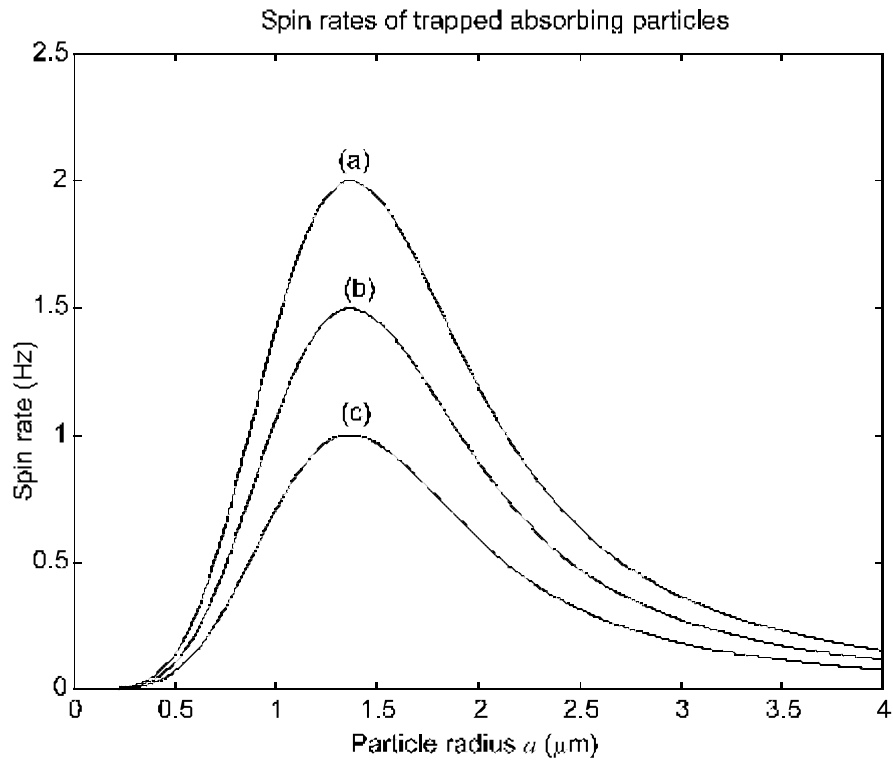

**Figure 4.** Spin rates of trapped particles. The spin rate of a particle trapped in the centre of the trap can be calculated from equation 18. Here, the particles are trapped in water by a 1 mW $LG_{03}$ laser beam with a width $w = 1$ $\mu$m in the trapping plane. The polarization of the beam affects the spin rate; case (a) is for a beam circularly polarized in the same direction as the orbital motion, (b) is for a plane-polarized beam, and (c) is for the case when the polarization and helicity are in opposite directions.

The particle will also absorb energy from the beam. A particle on the beam axis will absorb a total power of

$$P_a = \alpha P \left(1 - e^{\frac{-2a^2}{w^2(z)}} \sum_{k=0}^{l} \frac{1}{k!}\left(\frac{2a^2}{w^2(z)}\right)^k \right) \quad (19)$$

The temperature reached by the particle is of interest, as convective flow of the medium around the particle could exert a force on it. If damage to trapped particles is to be minimized, the temperature rise should be kept as low as possible. Under other circumstances, heating of the particle might be a desired effect.

Due to the small surface area of typical trapped particles, radiative loss of energy will be insignificant. As the length scales involved are very small, and convective fluid flow velocities will be small, the loss of heat through thermal conduction will predominate, and heat transport through convection will be insignificant[27].

For a spherical particle losing the heat absorbed from the beam by isotropic conduction into a surrounding medium of thermal conductivity $k$, the radial variation of the temperature $T$ in the surrounding medium is given by

$$T = \frac{P_a}{4\pi r k} + T_0$$

where $T_0$ is the background temperature of the medium. While most biological samples suffer only small temperature rises due to their relative transparency, strongly absorbing particles can experience very high temperatures. As shown in figure 5, large particles more readily lose heat by conduction, while small particles absorb little energy from the beam, so maximum rises are experienced by particles just large enough to absorb most of the beam power.

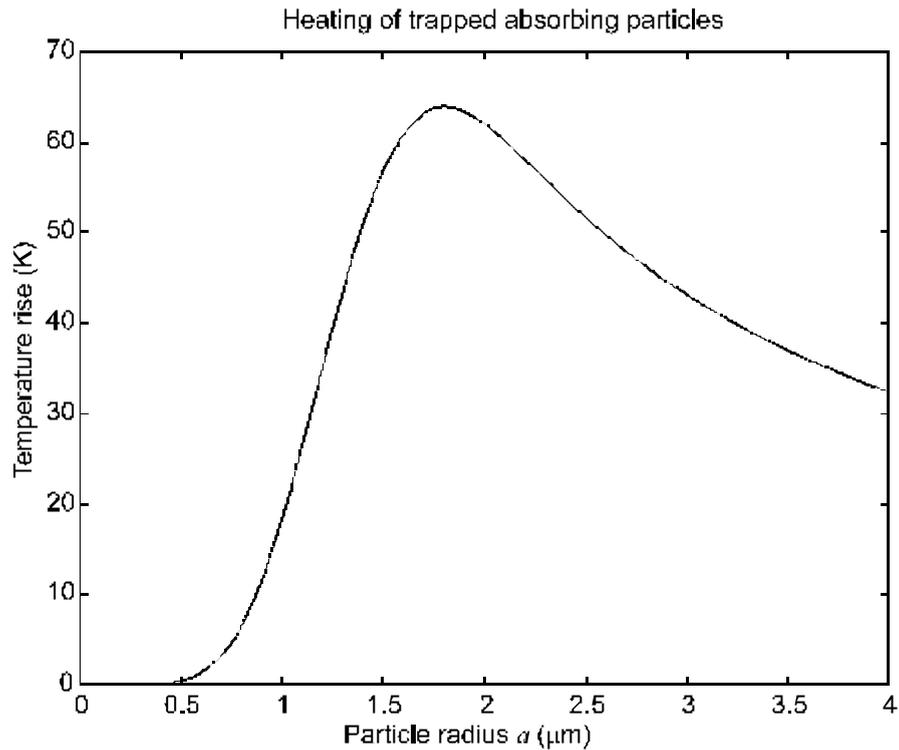

**Figure 5.** Temperature rise of completely absorbing trapped particles. The particles here are trapped in water by a 1 mW $LG_{03}$ laser beam of width 1 $\mu$m in the trapping plane. A large particle readily loses heat through conduction, while a small particle absorbs little energy from the beam. Particles of an intermediate size undergo the greatest heating.

## 3. Comparison with trapping in Gaussian laser beams

$TEM_{00}$ Gaussian laser beams are easily produced, and are often used for laser trapping and manipulation. Both Gaussian and LG beams can be used to trap either absorbing or transparent particles, but the resultant properties of the traps differ somewhat. In particular, a Gaussian beam does not carry orbital angular momentum. A comparison can show when the use of a doughnut beam trap should be preferred.

### 3.1 Transfer of orbital angular momentum

A Gaussian laser beam does not carry orbital angular momentum as the wavefronts are spherical rather than helical. As a result, a Gaussian beam can only transport angular momentum through circular polarization. If the application of a large torque to a trapped particle is desired, a doughnut beam would usually be employed. If a torque relatively independent of the position within the trap is wanted, a Gaussian beam should be used instead, as the torque depends on the intensity and polarization state only.

### 3.2    Difference in axial force and heating

Heating and axial forces are usually unwanted effects in optical traps, but can be quite large when absorbing particles are being trapped. Trapping in Laguerre-Gaussian beams allows these effects to be minimized, as the particle can be trapped in the dark central region of the beam. Particles will not always be on the beam axis, such as when they are moving into or out of the trap, and sufficiently large particles cannot be contained within the central low-intensity region, so the maximum forces experienced by particles as they move within the trap are of interest. The efficiency of an optical trap can be characterized by the ratio of the radial trapping force to the (unwanted) axial force. As the axial force is proportional to the absorbed power, the same efficiency can also be used to compare heating of particles within the trap. Figure 6 shows the ratio of the maximum radial trapping force to the unwanted maximum axial force for both $TEM_{00}$ and $LG_{03}$ beams under typical conditions.

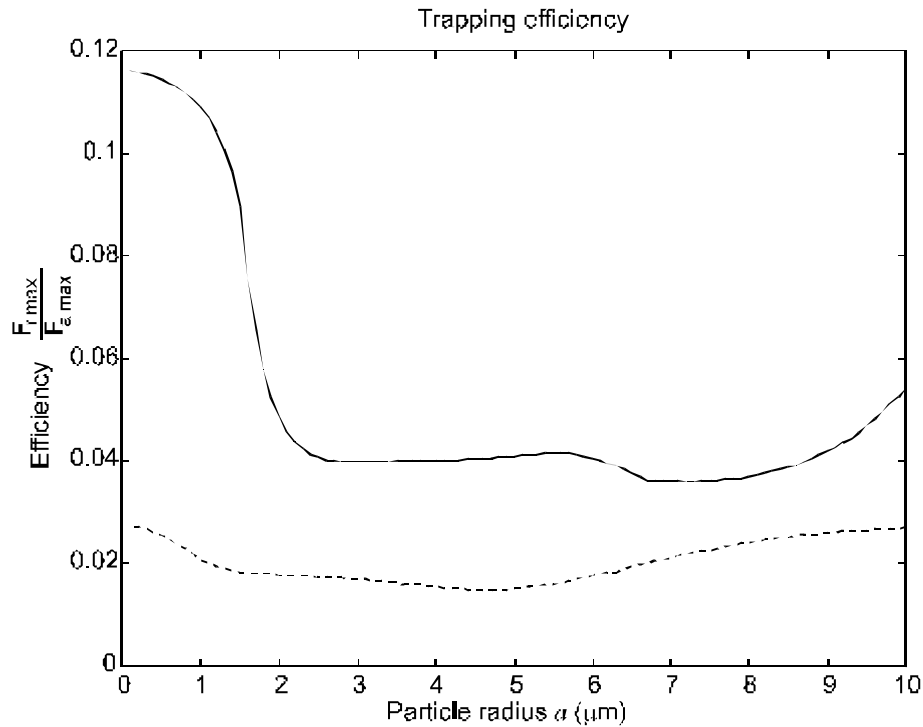

**Figure 6.** Trapping efficiency for absorbing particles. The efficiency of an absorbing particle optical trap can be measured by the ratio of the maximum radial trapping force to the (unwanted) maximum axial force. The beams are a $TEM_{00}$ Gaussian beam (dashed line) and an $LG_{03}$ doughnut beam (full line). Boths beams have a wavelength of 632.8 nm, and a waist width of 1 $m$m. The particles are trapped in water 5 $m$m before the beam waist. Traps using doughnut laser beams are generally more efficient, as the central region in a Gaussian beam provides a major part of the unwanted axial force, and only a minor contribution to the radial trapping. If a particle is sufficiently small, it can be contained entirely within the central dark region of the doughnut beam, and will experience very low axial forces, and a correspondingly higher efficiency will result.

The improvement obtained by using an LG beam is quite large. Axial forces will be much smaller, especially for small particles, and heating will be much less. Despite this, Gaussian beams are usually used in optical traps, as most optical traps are used to trap transparent particles where absorption effects are much smaller. Gaussian beams are also easier to produce and give higher intensity as losses in conversion to a doughnut beam are avoided, and the beam structure allows the power to be focussed into a smaller area.

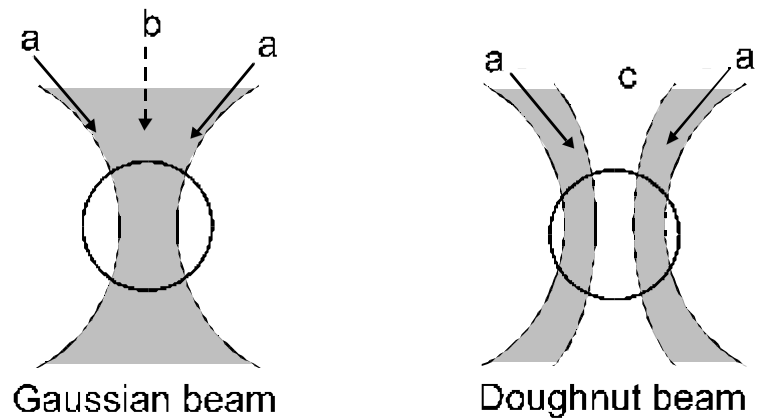

**Figure 7.** Heating of trapped transparent particles in Gaussian and doughnut beams. As rays incident on a transparent particle at large angles of convergence (a) are responsible for most of the trapping forces acting on the particle, removing the central rays (b) has relatively little effect on these forces. In a Gaussian beam, this central portion of the beam is the most intense, and is responsible for most of the heating and scattering forces acting on the particle. A Laguerre-Gaussian doughnut beam, however, has a very low intensity in the centre, with corresponding smaller scattering forces and heating. The axial trapping forces (and the axial spring constant) are greater for the doughnut beam trap as well, as a trapped particle encounters a large intensity gradient as it moves upwards or downwards into the low intensity region (c).

## 4. Multiple beam optical traps

Many applications of optical trapping use multiple laser beams. Although it is possible to trap transparent particles with a single beam, improved trapping can be obtained through using multiple beams, for example by using two counter-propagating beams to reduce total absorption forces while increasing gradient forces. Trapping can also be achieved without strongly convergent beams, as the intersection of two beams provides an intensity maximum to which high refractive index particles will be attracted. A single laser beam can trap absorbing particles only in two dimensions, but multiple beam traps suggest the possibility of trapping absorbing particles three-dimensionally.

A large number of different beam configurations can be used; a few basic configurations will be examined here to illustrate some simple effects that can be obtained by using multiple beams, with emphasis on the effects on absorbing particles (transparent particles exhibit relatively simple behaviour, high refractive index particles being attracted to the high-intensity regions of the beam). We can also note that while a trap using multiple mutually coherent beams is, properly speaking, a multiple beam trap, it can also be regarded as a single beam trap with a spatially complicated beam structure. Only the simpler case of incoherent beams will be dealt with here.

Most laser particle traps use a microscope to focus the beam onto a sample contained on a microscope slide on the stage. This imposes restrictions on the usable beam configurations, as beams can only be introduced from above and below the slide, ruling out traps using orthogonal beams and similar arrangements. An optical trap of different construction, possibly using optical fibres, can allow a greater variety of configurations. A few simple multiple beam traps are considered below. More complex multiple beam traps have been considered elsewhere[34]. We have not been able to identify any multiple beam configurations able to provide simple, effective, three dimensional trapping using a conventional microscope optical trap.

### 4.1 Confocal coaxial counter-propagating identical beams

One of the simplest cases is where two beams propagate in opposite directions along the same beam axis and share focal planes. If the beams are of the same width, power and wavelength, the radial and axial momenta carried by the beams will cancel, giving zero radial and axial absorption forces at all positions within the beam. This allows small gradient forces to be used to trap absorbing particles. (Polarization forces have been used to trap non-transparent particles in single beams[28].) The angular momenta of the two beams can either add or cancel, depending on the direction.

### 4.2 Coaxial counter-propagating beams

If the two coaxial beams do not share focal planes, absorption forces can be used to radially trap particles between the two focal planes if both beams are converging along a portion of the beam axis. The Poynting vector for such a configuration is shown in figure 8. The axial force acting on a small particle in the centre of the trap will be zero, but the equilibrium point is unstable axially. This configuration, however, does allow relatively strong radial trapping with minimal axial forces.

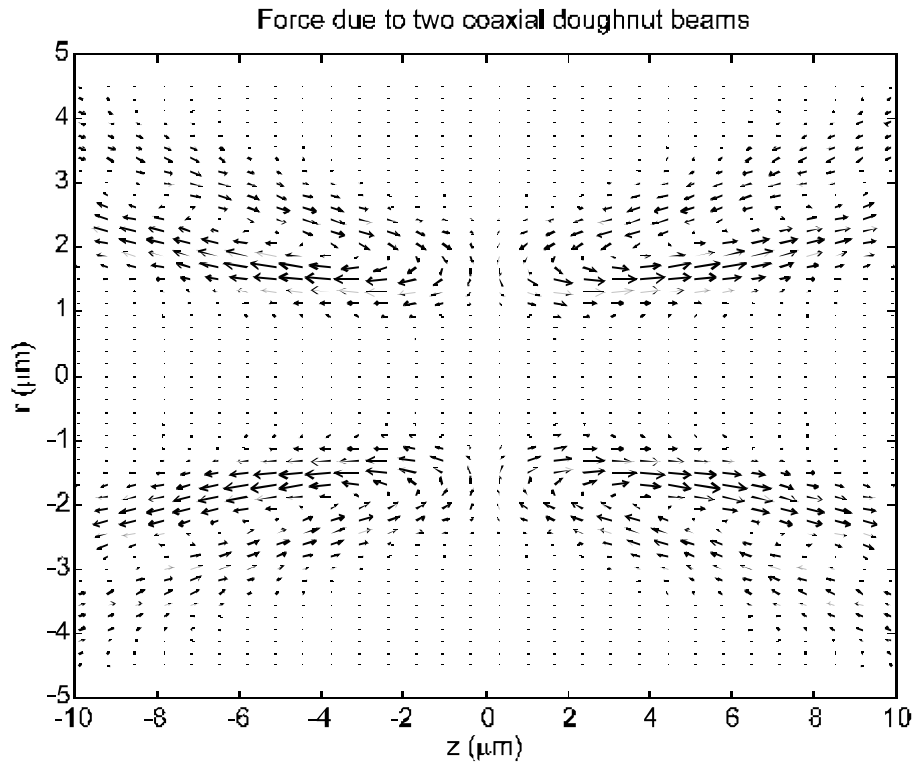

**Figure 8** Forces in counter-propagating coaxial doughnut trap. Identical doughnut laser beams propagating in opposite directions produce a trap with an equilibrium point in the centre. The equilibrium is axially unstable, so true three-dimensional trapping is not achieved. A very low axial force is present near the centre of the trap. The length of the arrows on the diagram are proportional to the magnitude of the force. The beams are both 632.8 nm $LG_{03}$ beams with waist widths of 1 *m*m, with the focal planes 2 *m*m apart.

## 5. Orbital Motion

If a particle illuminated by a Laguerre-Gaussian laser beam absorbs any momentum from the beam, it will also gain angular momentum about the beam axis through absorption of the orbital angular momentum of the beam. A particle trapped on the beam axis (the location of the phase singularity) will spin. If the particle is off the beam axis, orbital motion about the beam axis will result. This orbital motion will not be directly affected by the angular momentum carried by any circular polarization of the beam.

Of interest is under what conditions a particle undergoing such orbital motion will move into the centre of the trap, escape, or remain in a stable orbit.

### 5.1 Escape through orbital motion

Obviously, an orbiting particle (other than an atom, which comprises a special case considered below) in a vacuum will not remain trapped as its tangential speed will continue to increase. The cases of interest involve particles moving in a viscous medium. The Reynolds numbers expected in typical optical traps are very small (Re $\ll 1\times10^{-5}$), so the motion will be dominated by viscous forces, flow will be laminar, and the particle velocity will be close to the terminal velocity within the fluid.

For low Reynolds number motion, the drag force acting on a particle can be expressed simply. For a sphere of radius *a* moving uniformly with a velocity **u** through a fluid of viscosity $\eta$, the drag force is given by Stokes Law

$$\mathbf{D} = -6\pi\eta a\mathbf{u}. \qquad (21)$$

For other shapes, similar results with slightly different numerical coefficients will apply. The dependence of the drag force on shape is small, and the drag force acting on a sphere (equation 21) will be assumed in all cases. The dependence on surface texture is negligible. It should be noted that the drag force can be affected by a nearby surface[35].

A spherical particle of radius *a* and mass *m* experiencing an external force **F** will have a terminal velocity given by

$$\dot{\mathbf{x}} = \mathbf{F}/6\pi\eta a. \qquad (22)$$

In the case here, the force **F** is the optical force acting on the particle. The optical force is a function of position within the trap and is independent of time and particle velocity, so the resultant equation of motion for the particle is

$$\mathbf{x}(t) = \frac{\int_0^t \mathbf{F}(\mathbf{x}(t))dt}{6\pi\eta a} + \mathbf{x}(0). \qquad (23)$$

Since we are primarily interested in whether or not the particle will escape, move in a stable orbit, or be forced into the centre of the trap (i.e. onto the beam axis), we can usefully express this in cylindrical coordinates

$$\dot{r} = \frac{F_r(r,\boldsymbol{f},z) + \frac{m}{r}\left(\frac{F_r(r,\boldsymbol{f},z)}{6\boldsymbol{p}h a}\right)^2}{6\boldsymbol{p}h a}$$

$$\dot{\boldsymbol{f}} = \frac{F_f(r,\boldsymbol{f},z)}{6\boldsymbol{p}h a} \qquad (24)$$

$$\dot{z} = \frac{F_z(r,\boldsymbol{f},z)}{6\boldsymbol{p}h a}$$

In particular, if $\dot{r} = 0$, the particle will be in a closed orbit, if $\dot{r} < 0$, the particle will move towards the beam axis, and if $\dot{r} > 0$, the particle will move outwards. Thus, it is sufficient to consider the sign of the expression

$$F_r(r,\boldsymbol{f},z) + \frac{m}{r}\left(\frac{F_r(r,\boldsymbol{f},z)}{6\boldsymbol{p}h a}\right)^2. \qquad (25)$$

### 5.2   Orbital motion of absorbing particles

If the particle is constrained to move in a fixed z plane, we need consider radial and azimuthal forces only. The optical forces acting on a small spherical particle of radius $a$ are

$$F_r = \frac{\boldsymbol{p}a^2}{c}S_r = \frac{2p!\,Pa^2}{c(l+p)!}\left(\frac{2r^2}{w^2}\right)^l L_p^{2l}\!\left(\frac{2r^2}{w^2}\right)\frac{1}{w^2}e^{\frac{-2r^2}{w^2}}\frac{zr}{z_r^{\,2}+z^2}$$

$$F_f = \frac{\boldsymbol{p}a^2}{c}S_f = \frac{2p!\,Pa^2}{c(l+p)!}\left(\frac{2r^2}{w^2}\right)^l L_p^{2l}\!\left(\frac{2r^2}{w^2}\right)\frac{1}{w^2}e^{\frac{-2r^2}{w^2}}\frac{l}{kr} \qquad (26)$$

For the cases of interest here ($p = 0$), this gives

$$F_r = \frac{2Pa^2}{c\,l!}\left(\frac{2r^2}{w^2}\right)^l \frac{1}{w^2}e^{\frac{-2r^2}{w^2}}\frac{zr}{z_r^{\,2}+z^2}$$

$$F_f = \frac{2Pa^2}{c\,l!}\left(\frac{2r^2}{w^2}\right)^l \frac{1}{w^2}e^{\frac{-2r^2}{w^2}}\frac{l}{kr} \qquad (27)$$

with expression 25 becoming

$$\frac{2Pa^2}{c\,l!}\left(\frac{2r^2}{w^2}\right)^l\frac{1}{w^2}e^{\frac{-2r^2}{w^2}}\frac{zr}{z_r^{\,2}+z^2} + \frac{m}{36\boldsymbol{p}^2\boldsymbol{h}^2 a^2 r}\left(\frac{2Pa^2}{c\,l!}\left(\frac{2r^2}{w^2}\right)^l\frac{1}{w^2}e^{\frac{-2r^2}{w^2}}\frac{l}{kr}\right)^2 \qquad (28)$$

or, equivalently,

$$\frac{w^2 z r^4}{w_0^2} + \frac{a^3 \rho P l^2}{54 \rho h^2 c l!} \left(\frac{2r^2}{w^2}\right)^l \frac{1}{w^2} e^{\frac{-2r^2}{w^2}} \tag{29}$$

where $\rho$ is the density of the particle. The value of this expression depends on the properties of the fluid, the beam, and the particle. This expression, of the general form

$$r^{2l-2} e^{-hr^2} - g \tag{30}$$

where $g$ and $h$ are positive (since $z < 0$), is equal to zero for

$$r = g^{\frac{1}{2(l-1)}} \exp\left(-\frac{1}{2} W\left(\frac{-h}{l-1} g^{\frac{1}{l-1}}\right)\right) \tag{31}$$

where $W$ is Lambert's W function. The expressions $g$ and $h$ are equal to

$$g = \frac{-54 \rho h^2 c l! z w^2}{a^3 \rho P l^2 w_0^2} \left(\frac{w^2}{2}\right)^l$$

$$h = \frac{2}{w^2} \tag{32}$$

giving for a typical case for small particles ($P = 10^{-3}$ W, $l = 3$, $\eta = 10^{-3}$ Nsm$^{-2}$, c = $2.25 \times 10^8$ ms$^{-1}$, $z = -10^{-6}$ m, $w = 10^{-6}$ m, $w_0 = 0.5 \times 10^{-6}$ m, $a = 0.1 \times 10^{-6}$ m and $\rho = 2000$ kgm$^{-3}$) $g = 6.36 \times 10^{-18}$ and $h = 2 \times 10^{12}$. Under these conditions ($|x| \gg 1$, $x < 0$), $W(x)$ does not exist, and no radius $r$ corresponds to an orbit. All particles will, in principle, become trapped. For large radii, the optical force will be very small, and thermal motion will predominate.

With fluids of much lower viscosity, such as gases, an orbit radius can be found. Under these conditions ($|x| \ll 1$), $W(x) \approx x$, so the orbit radius becomes

$$r = g^{\frac{1}{2(l-1)}}. \tag{33}$$

This orbit is not stable, and indicates the radius within which particles will become trapped. This radius of "trapability" can be comparable to the beam radius, providing a limit to the effective size of the trap.

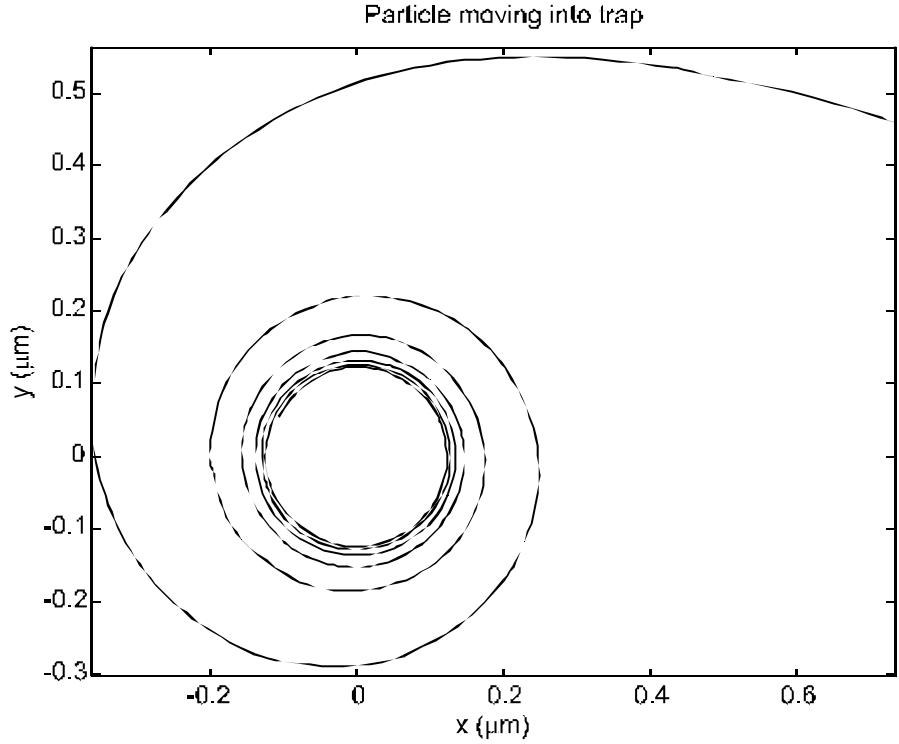

**Figure 9**. A particle moving into an optical vortex trap. This numerical simulation shows a particle moving into the centre of an optical trap under the conditions described above. The particle begins at rest 5 *m*m from the centre of the trap. The particle moves inwards without orbiting until it reaches the bright ring of the doughnut beam, at which time it begins to orbit the centre of the trap, spiralling inwards. The simulation is carried out until the particle is within a particle radius from the centre of the trap and the assumptions made cease to be valid.

### 5.3 Orbital motion of small transparent particles

Small transparent particles will be trapped in the high intensity portion of the beam (see figure 1) rather than on the beam axis. If the particles are slightly absorbing, they will move in orbits staying within the intense portion of the beam (strongly absorbing particles will not be trapped). Unless the fluid within the trap has a very low viscosity, the speeds will not become high enough to cause the particle to escape.

The radial trapping force acting on a small particle of radius $a$ and refractive index $n$ in a medium of refractive index $n_m$ is

$$\begin{aligned} F_r &= \frac{2\pi a^3 \varepsilon_0}{3}\left(\frac{1}{n_m^2} - \frac{1}{n^2}\right)\frac{\partial}{\partial r} E_0^2 \\ &= \frac{2\pi a^3 \varepsilon_0}{3}\left(\frac{1}{n_m^2} - \frac{1}{n^2}\right) E_0^2 \frac{2}{r}\left(l - \frac{2r^2}{w^2}\right) \end{aligned} \quad (34)$$

giving an equilibrium radius of

$$r_0 = w_0 \sqrt{\frac{l}{2}}. \tag{35}$$

The azimuthal absorption force acting on the particle is

$$F_f = \frac{\sigma}{c} S_f = \frac{2P\sigma}{c\pi l!} \left(\frac{2r^2}{w^2}\right)^l \frac{1}{w^2} e^{\frac{-2r^2}{w^2}} \frac{l}{kr} \tag{36}$$

where the absorption cross-section $\sigma$ is given in terms of the microscopic extinction coefficient $\lambda$ by the volume times $\lambda$ when the particle is only slightly absorbing, giving for a spherical particle $\sigma = 4/3 \, \pi a^3 \lambda$. In most cases, the particle will remain near the equilibrium radius $r_0$ (otherwise, the procedure as used for absorbing particles can be carried out), giving

$$F_f = \frac{2P\sigma}{c\pi l!} l^l e^{-l} \frac{1}{w^2} \frac{\sqrt{2l}}{kw_0^3}. \tag{37}$$

The major point of interest is the orbital speed $u$ and frequency $f$, which will be given by

$$u = \frac{\sqrt{2}\sigma P l^{l+\frac{1}{2}} e^{-l}}{3\pi^2 \eta a c k l! w_0^3} \tag{38}$$

and

$$f = \frac{\sigma P l^l e^{-l}}{3\pi^3 \eta a c k l! w_0^4}. \tag{39}$$

For a typical case, such as a sphere trapped in water ($P = 10 \times 10^{-3}$ W, $l = 3$, $\eta = 10^{-3}$ Nsm$^{-2}$, $c = 2.25 \times 10^8$ ms$^{-1}$, $k = 1.32 \times 10^7$ m$^{-1}$, $w_0 = 0.5 \times 10^{-6}$ m, $a = 0.1 \times 10^{-6}$ m and $\lambda = 7.5 \times 10^4$ m$^{-1}$), this gives $u = 1.57 \times 10^{-6}$ ms$^{-1}$ and $f = 0.407$ Hz. In this case, with a viscous medium and the particle absorbing 1% of the incident light, the orbital speed is slow. The orbital speed is proportional to the beam power and particle size (although the approximations employed here will cease to be valid for large particles) so a judicious choice of trap and particle can give observable motion. As optically trapped particles are typically very small, even a relatively large microscopic extinction coefficient will only result in moderate absorption. Trapping in a less viscous medium will give higher orbital speeds, and the possibility of escape will need to be considered.

## 6. Trapping and cooling of atoms

Atoms in laser beams show behaviour similar to both transparent and absorbing particles, experiencing gradient and absorption forces. The behaviour of atoms also has some unique features, such as the velocity dependence of forces due to the strong wavelength dependence of the interaction with the incident field and Doppler shifts.

A simple semi-classical analysis shows the major features of the interaction between the atom and the laser field. A two level atom, with upper and lower level occupation probabilities $P_u$ and $P_l$ and resonant frequency $w_0$ will experience two types of absorption/emission interactions with the field, namely spontaneous emission and stimulated emission and absorption. An average force will result from the occasions on which the atom absorbs a photon from the incident field and undergoes spontaneous emission (since the average momentum transfer due to spontaneous emission is zero). This force is given by

$$\langle \mathbf{F}_a \rangle = \hbar \mathbf{k} \Gamma P_u \qquad (40)$$

where $\mathbf{k}$ is the wave-vector of the laser field and $\Gamma$ is the spontaneous emission rate. The upper state occupation $P_u$ depends on the detuning $\Delta = (w - w_0 + d)$, where $w$ is the laser frequency and $d$ is the Doppler shift. It also depends on the laser intensity and the line width, which is dependent in turn on the spontaneous emission rate and the laser intensity. The upper state occupation can be simply calculated for weak fields, where it is proportional to the line strength and laser intensity, and for strong fields, where $P_u = \frac{1}{2}$. An identical result can be obtained using a quantum mechanical approach[36].

Since a doughnut beam carries angular momentum of $l\hbar$ per photon, there will be a torque about the beam axis, given by

$$\langle \mathbf{\tau} \rangle = l\hbar \Gamma P_u \qquad (41)$$

The electric field of the laser light will induce a polarization in the atom of $\mathbf{d} = \mathbf{a}\mathbf{E}$, where $\mathbf{a}$ is the polarizability of the atom (which can be calculated quantum mechanically), giving an interaction energy of

$$U = -\mathbf{E} \cdot \mathbf{d} = -\mathbf{a}E^2. \qquad (42)$$

This will result in a time-averaged force of

$$\mathbf{F}_g = \frac{\mathbf{a}}{2} \nabla E^2. \qquad (43)$$

The polarizability **a** depends on the detuning, and will result in the atoms being either attracted to or repelled away from high intensity regions of the beam. The similarities between the effects of laser beams on transparent and absorbing particles and atoms can be readily seen. The major difference is in the dependence of forces and torques on the detuning and Doppler shifts. Closely related is the question of whether heating or cooling of trapped atoms occurs, as heating can result in atoms gaining sufficient energy to escape from the trap.

We can also note that the interaction of the atom with the field is a discrete and probabilistic process. For a high-intensity incident field and a strongly interacting atom, it can be approximated as a continuous process and calculations of trajectories performed, but this will not be possible for weak fields or weak interactions. In these cases, sample paths can be calculated using Monte-Carlo methods, still using the assumption that the motion of the atom can be treated classically.

Apart from the possibility of optical trapping, the Doppler shifts experienced by moving atoms can be utilized to cool an atomic gas to such low temperatures that they can be confined by a weak magnetic trap (MOT).

In view of the great success of optical trapping and cooling of atoms and ions in recent years it is interesting to speculate on what role optical vortices might play in that area. In a series of papers, Allen and coworkers have studied the effect on individual atoms and ions of LG beams containing central vortices. They showed[36,37] that two level atoms experienced an additional Doppler shift on account of the helical phase structure and also an azimuthal force which applies a torque about the beam axis. In the saturation limit this is given by

$$\left|\langle \boldsymbol{t}_D \rangle\right| \approx l\hbar \frac{\Gamma}{2}. \tag{44}$$

This corresponds to transfer of angular momentum absorbed from the beam to the atom while the energy absorbed is reradiated away, isotropically on the average. These results are consistent with our
simpler model of absorbing particles.

They also showed[37] that if the atom (or ion) is trapped in a potential well the result can be a spiralling out to a constant orbit. A further case considered was an ion undergoing cyclotron motion[38] in a magnetic field irradiated by counter-propagating LG beams. Depending on the helicity of the beams, they predicted that the cyclotron motion could be enhanced or damped away with the atom spiralling in to rest on the axis.

A few reports of atom trapping experiments using LG beams have been published. One attraction of the use of the vortex is the possibility of using a gradient force to trap an atom on the axis, in the dark, where excitation and heating are minimized. This could be achieved by detuning the laser to the anomalous dispersion side of the spectral line so that the atom behaves like a particle with refractive index less than unity. Unfortunately this is the blue side of the spectral line, while Doppler cooling requires illumination on the red side of the line so that it does not seem possible to use the same beam to simultaneously cool and trap.

A doughnut mode has been used with a magneto-optic trap by Snadden et al.[39]. They used an intense $LG_{01}$ beam tuned to the red to cool a large volume of vapour surrounding a central MOT using weaker $TEM_{00}$ beams for trapping. They point out that high intensity is needed for efficient cooling but is not needed to trap atoms which have already been cooled. The low intensity in the trap also results in minimal AC Stark shifts and broadening, which is useful if high-precision measurements are to be made. In previous work they had used beams with central minima produced by transmitting $TEM_{00}$ beams through glass plates with central spots, but the use of vortex beams avoids the tendency of the dark spot to fill by diffraction as the beam propagates.

Experiments have also been reported by Schiffer et al.[40] involving the focussing of a beam of cold Ne atoms by a coaxial $TEM^*_{01}$ doughnut produced by a computer generated hologram. This group has also proposed a dipole force trap based on vortex beams which can be loaded with precooled atoms. Recently, Kuga et al.[41] reported a trap for Rubidium atoms based on a $LG_{03}$ doughnut beam, retaining the atoms for around 1.5 s, the losses being dominated by background gas collisions. Excess heating due to the trapping beam is prevented by the use of pulsed polarisation gradient cooling.

Yin et al.[42] have also shown that a doughnut beam can be used to guide cold atoms from a MOT. With a very weak repumping beam, the doughnut beam can be converted into a gravito-optical trap (GOT), in which Sisyphus cooling and geometrical cooling can reduce the temperature to about 1 *m*K.

# 7.   Conclusion

Optical vortices exert mechanical forces which can be easily detected and which can even be useful in the context of trapping microscopic particles and atoms. Doughnut beams can be used for trapping high refractive index particles with higher efficiency and less heating, and can be used to trap low refractive index particles and bubbles.

When absorbing particles are trapped in a doughnut beam ,which is easier than in a Gaussian beam, angular momentum is transferred giving rise to rotation and perhaps some orbital motion. Doughnut beams are also finding increasing use in atom and ion traps.

# References

[1] M. Berry Singularities in waves and rays in R. Balian, M. Kleman and J.-P. Poirier, eds "Physics of Defects" North-Holland, Amsterdam (1981)

[2] P.N. Lebedev Untersuchungen über die Druckkräfte des Lichtes *Annalen der Physik* **6**, 433 (1901)

[3] E.F. Nichols and G.F. Hull A preliminary communication on the pressure of heat and light radiation *Physical Review* **13**, 307 (1901)

[4] R.A. Beth Mechanical detection and measurement of the angular momentum of light *Physical Review* **50**, 115-125 (1936)

[5] M.E.J. Friese, T.A. Nieminen, N.R. Heckenberg and H. Rubinsztein-Dunlop Optical alignment and spinning of laser-trapped microscopic particles *Nature* **394**, 348-350 (1998) (note that figure 1 is incorrectly printed)

[6] L. Allen, M.W. Beijersbergen, R.J.C. Spreeuw and J.P. Woerdman Orbital angular momentum of light and the transformation of Laguerre-Gaussian laser modes *Physical Review A* **45**, 8185-8188 (1992)

[7] E. Abramochkin and V. Volostnikov Beam transformations and nontransformed beams *Optics Communications* **83**, 123-135 (1991)

[8] C. Tamm and C.O. Weiss Bistability and optical switching of spatial patterns in a laser beam *Journal of the Optical Society of America B - Optical Physics* **7**, 1034-1038 (1990)

[9] K. Dholakia, N.B. Simpson, M.J. Padgett and L. Allen Second-harmonic generation and the orbital angular momentum of light *Physical Review A* **54**, R3742-R3745 (1996)

[10] M. Padgett, J. Arlt, N. Simpson and L. Allen An experiment to observe the intensity and phase structure of Laguerre-Gaussian laser modes *American Journal of Physics* **64**, 77-82 (1996)

[11] M.W. Beijersbergen, L. Allen, H.E.L.O. van der Veen and J.P. Woerdman Astigmatic laser mode converters and transfer of orbital angular momentum *Optics Communications* **96**, 123-132 (1993)

[12] S.M. Barnett and L. Allen Orbital angular momentum and nonparaxial light beams *Optics Communications* **110**, 670-678 (1994)

[13] I.V. Basistiy, V.Y. Bazhenov, M.S. Soskin and M.V. Vasnetsov Optics of light beams with screw dislocations *Optics Communications* **103**, 422-428 (1993)

[14] A. Ashkin and J.M. Dziedzic Stability of optical levitation by radiation pressure *Applied Physics Letters* **24**, 586-588 (1974)

[15] A. Ashkin, J.M. Dziedzic, J.E. Bjorkholm and S. Chu Observation of a single-beam gradient force optical trap for dielectric particles *Optics Letters* **11**, 288-290 (1986)

[16] W.H. Wright, G.J. Sonek and M.W. Berns Parametric study of the forces on microspheres held by optical tweezers *Applied Optics* **33**, 1735-1748 (1994)

[17] W.H. Wright, G.J. Sonek, Y. Tadir and M.W. Berns Laser trapping in cell biology *IEEE Journal of Quantum Electronics* **26**, 2148-2157 (1990)

[18] A. Ashkin Forces of a single-beam gradient laser trap on a dielectric sphere in the ray optics regime *Biophysical Journal* **61**, 568-581 (1992)

[19] M.E.J. Friese, H. Rubinsztein-Dunlop, N.R. Heckenberg and E.W. Dearden Determination of the force constant of a single-beam gradient trap by measurement of backscattered light *Applied Optics* **35**, 7112-7117 (1996)

[20] N.B. Simpson, K. Dholakia, L. Allen and M.J. Padgett Mechanical equivalence of spin and orbital angular momentum of light - an optical spanner *Optics Letters* **22**, 52-54 (1997)

[21] N.B. Simpson, L. Allen and M.J. Padgett Optical tweezers with increased trapping efficiency *CLEO '97* CTuD7 (1997)

[22] G. Roosen and C. Imbert The $TEM^*_{01}$ mode laser beam - a powerful tool for optical levitation of various types of spheres *Optics Communications* **26**, 432-436 (1978)

[23] K. Sasaki, M. Koshioka, N. Kitamura and H. Masuhara Optical trapping of a metal particle and a water droplet by a scanning laser beam *Applied Physics Letters* **60**, 807-809 (1992)

[24] H. He, N.R. Heckenberg and H. Rubinsztein-Dunlop Optical particle trapping with higher-order doughnut beams produced using high efficiency computer generated holograms *Journal of Modern Optics* **42**, 217-223 (1995)

[25] S. Sato, Y. Harada and Y. Waseda Optical trapping of microscopic metal particles *Optics Letters* **19**, 1807-1809 (1994)

[26] M.E.J. Friese, T.A. Nieminen, N.R. Heckenberg and H. Rubinsztein-Dunlop Optical torque controlled by elliptical polarization *Optics Letters* **23**, 1-3, (1998)

[27] H. Rubinsztein-Dunlop, T.A. Nieminen, M.E.J. Friese and N.R. Heckenberg Optical trapping of absorbing particles *Advances in Quantum Chemistry* **30**, 469-492 (1998)

[28] K. Svoboda and S.M. Block Optical trapping of metallic Rayleigh particles *Optics Letters* **19**, 930-932 (1994)

[29] B.T. Unger and P.L. Marston Optical levitation of bubbles in water by the radiation pressure of a laser beam: An acoustically quiet levitator *Journal of the Acoustical Society of America* **83**, 970-975 (1988)

[30] K.T. Gahagan and G.A. Swartzlander Optical vortex trapping of particles *Optics Letters* **21**, 827-829 (1996)

[31] H. He, M.E.J. Friese, N.R. Heckenberg and H. Rubinsztein-Dunlop Direct observation of transfer of angular momentum to absorptive particles from a laser beam with a phase singularity *Physical Review Letters* **75**, 826-829 (1995)

[32] M.E.J. Friese, H. He, N.R. Heckenberg and H. Rubinsztein-Dunlop Transfer of angular momentum to absorbing particles from a laser beam with a phase singularity *Proceedings of SPIE* **2792**, 190-195 (1996)

[33] M.E.J. Friese, J. Enger, H. Rubinsztein-Dunlop and N.R. Heckenberg Optical angular-momentum transfer to trapped absorbing particles *Physical Review A* **54**, 1593-1596 (1996)

[34] N.R. Heckenberg, T.A. Nieminen, M.E.J Friese and H. Rubinsztein-Dunlop Trapping microscopic particles with singular beams *Proceedings of SPIE* **3487**, 46-53 (1998)

[35] J. Happel and H. Brenner "Low Reynolds number hydrodynamics" 2nd rev. ed., Noordhoff International Publishing, Leiden (1973)

[36] M. Babiker, W.L. Power and L. Allen Light-induced torque on moving atoms *Physical Review Letters* **73**, 1239-1241 (1994)

[37] W.L. Power and R.C. Thompson Laguerre-Gaussian laser beams and ion traps *Optics Communications* **132**, 371-378 (1996)

[38] M. Babiker, V.E. Lembessis, W.K. Lai and L. Allen Doppler cooling of ion cyclotron motion in counter-propagating Laguerre-Gaussian beams *Optics Communications* **123**, 523-529 (1996)

[39] M.J. Snadden, A.S. Bell, R.B.M. Clarke, E. Riis and D.H. McIntyre Doughnut mode magneto-optical trap *Journal of the Optical Society of America B -Optical Physics* **14**, 544-552 (1997)

[40] M. Schiffer, G. Wokurka, M. Rauner, T. Slawinski, M. Zinner, K. Sengstock and W. Ertmer Holographically designed light-fields as elements for atom optics *IQEC '96 Sydney 14th July 1996 Proceedings*, MC3, 1-2 (1996)

[41] Y. Torii, N. Shiokawa, T. Hirano, T. Kuga, Y. Shimizu and H. Sasada Pulsed polarization gradient cooling in an optical dipole trap with a Laguerre-Gaussian laser beam *European Physical Journal D* **1**, 239-242 (1998)

[42] J.P. Yin, Y.F. Zhu, W.B. Wang, Y.Z. Wang and W. Jhe Optical potential for atom guidance in a dark hollow laser beam *Journal of the Optical Society of America B - Optical Physics* **15**, 25-33 (1998)